\documentclass[aps,preprint,pdftex,showpacs]{revtex4}
\begin{document}
\draft
\title{Coherent Control of Resonance-Mediated Reactions:  F + HD}
\author{Vlado Zeman, Moshe Shapiro\cite{bylineone} and Paul Brumer}
\affiliation{Chemical Physics Theory Group, Department of Chemistry, University
of Toronto, Toronto, Canada M5S 3H6\\}
\date{\today}
\begin{abstract}

Cross sections resulting from scattering that proceeds via an intermediate
resonance are shown to be exceptionally controllable using a coherent
superposition of only two initial states. Full quantum computations on
$\mathrm{F+HD}(v=0;j=0,1) \rightarrow \mathrm{H+ DF}, \mathrm{D+HF}$,
which exhibits a resonance in one of the
reactive channels, support the formal arguments, showing that control is
indeed vast. In this case the ratio of reactive integral cross sections
can be altered by a factor of 62 (compared to a non-coherent factor of
only 3.3), while the ratio of reactive differential cross sections can be
altered by a factor of over 6000 (compared to a non-coherent factor of
less than 7). These results constitute the first prediction of extensive
quantum control in a collisional process.
\end{abstract}
\pacs{34.50.Lf, 34.50.Rk}
\maketitle

Coherent control  is a rapidly developing method for controlling atomic
and molecular processes.  Recent experimental and theoretical developments
in this area have been extensively summarized in Refs. \cite{CC1,CC2}. In
this approach one manipulates, through laboratory parameters, quantum
interference contributions that arise when a final state is reached by two
or more coherent routes. Although initially introduced for unimolecular
processes, coherent control has been extended formally to collisions
\cite{Shapiro96,Brumer99,Frishman99a,Gong03} and has been numerically
studied for reactive scattering of the isotopic variants of
$\mathrm{H+H_2}$ \cite{Shapiro96} and for $\mathrm{H_2+H_2}$
\cite{Gong03}. All of the established principles apply equally well to
atomic, molecular and nuclear processes.

There are very few general results known for control of scattering. We
previously showed that complete control, i.e. the ability to completely
suppress or maximally enhance a particular product arrangement channel, is
achieved in collision processes when the number of reactant states used in
an initial coherent superposition is greater than the number of open final
product states in that product channel \cite{Frishman99a}.  In this
article we show that final product channels that arise via the formation
of a resonance can be completely controlled via a superposition of {\em
only two initial states, regardless of the number of open product states}.
We provide, as a numerical example, results of a full 3D quantum
computation of control through the recently discovered reactive resonance
in $\mathrm{F+HD} \rightarrow
\rm{D+HF}$\cite{Castillo98,Skodje00,Schatz00}. Even though this resonance
is short-lived \cite{Skodje00}, and direct reactive scattering occurs as
well, we find that a very large range of control is possible.

Control of a collisional process is achieved\cite{Shapiro96} by
constructing an initial state $|\Psi\rangle$ in the initial arrangement
channel composed of a superposition of $N$ energetically degenerate
asymptotic states $|q,m\rangle$.  Here $q$ denotes the arrangement channel
and $m$ indicates all state labels;  the energy label is dropped for
convenience. For the case of $N=2$:
\begin{equation}
    |\Psi\rangle = c_1|q,1\rangle + c_2|q,2\rangle \;.
\end{equation}
The cross section $\sigma_{q'}$ for scattering into final arrangement
channel $q'$, starting from state $|\Psi\rangle$, is given by
\begin{equation}
   \sigma_{q'} = \sum_n \left| \sum_{i=1}^2 c_i \langle q',n | T | q,i \rangle
      \right|^2 \;,
\label{xs1}
\end{equation}
where $T$ is the transition operator. For notational convenience we drop
the $q'$ subscript unless necessary.

Equation (\ref{xs1}) can be rewritten as
\begin{equation}
    \sigma_{q'} =
      |c_1|^2\sigma_{q'}(11) + |c_2|^2\sigma_{q'}(22)
      + 2{\mathrm Re}\left[ c_1^*c_2\sigma_{q'}(12) \right] \;,
\label{xs2}
\end{equation}
where
\begin{equation}
    \sigma_{q'}(ij) = \sum_n \langle q,i |T| q',n \rangle \langle q',n |T|
       q,j \rangle \;.
\label{xs3}
\end{equation}
To produce the integral cross section, the sum over $n$ includes an
integral over the product scattering angles as well as
a sum over all final diatom states.

Equation (\ref{xs2}) can be rewritten by defining the relative amplitude
    $s = { |c_2|^2 / [|c_1|^2 + |c_2|^2] }, 0 \le s \le 1 \;,$
and the relative phase between the two coefficients as $\phi_{12} =
\mathrm{Arg}\left[ c_2/c_1 \right]$. The cross section then becomes
\begin{eqnarray}
    \sigma_{q'} & = &
      (1-s)\sigma_{q'}(11) + s\sigma_{q'}(22) \nonumber \\
      & \!\!\!\!\!\!\!\!\!\!\!\!\!\!\!\! + & \!\!\!\!\!\!\!\!
      2\sqrt{s(1-s)}\left| \sigma_{q'}(12) \right|
      \cos\left( \mathrm{Arg}[\sigma_{q'}(12)] + \phi_{12} \right) \;,
\label{sigma}
\end{eqnarray}
indicating that control is achieved experimentally by varying $s$ and
$\phi_{12}$. Note that $\sigma_{q'}(11)$ and $\sigma_{q'}(22)$ are real,
corresponding to the cross sections for initial states 1 and 2
respectively, while $\sigma_{q'}(12)$ is complex, corresponding to quantum
interference between the two reaction pathways. Since $\phi_{12}$ can
always be chosen so that $\mathrm{Arg}[\sigma_{q'}(12)] + \phi_{12}$ is a
multiple of $2\pi$, the extent of control over $\sigma_{q'}$ is determined
by $|\sigma_{q'}(12)|$, which is seen to satisfy the Schwartz inequality $
|\sigma_{q'}(12)| \le \sqrt{ \sigma_{q'}(11)\sigma_{q'}(22) }$. Maximum
control therefore occurs when $|\sigma_{q'}(12)| = \sqrt{
\sigma_{q'}(11)\sigma_{q'}(22)}$.  The value of the cross section can be
controlled within the range $\sigma_{q'}(\min) \le \sigma_{q'} \le
\sigma_{q'}(\max)$, where the limiting values and corresponding control
parameters are [assuming $\sigma_{q'}(22) > \sigma_{q'}(11)$]
$$
  \begin{array}{rlrl}
    \sigma_{q'}(\min) =  & 0 &
    \sigma_{q'}(\max) = & \sigma_{q'}(11) + \sigma_{q'}(22) \\
    s(\min) = & { \sigma_{q'}(11) \over \sigma_{q'}(11) + \sigma_{q'}(22) } &
    s(\max) =  & { \sigma_{q'}(22) \over \sigma_{q'}(11) + \sigma_{q'}(22) } \\
    \phi_{12}(\min) = & \pi - \mathrm{Arg}\left[ \sigma_{q'}(12) \right]
    & \phi_{12}(\max) = & - \mathrm{Arg}\left[ \sigma_{q'}(12)
    \right] \;. \end{array}
$$
Note that $\sigma_{q'}(\max)$ is twice that achievable in the absence of
the interference term.

We show below that the Schwartz equality is satisfied for scattering into
a particular arrangement channel if it occurs through a resonance at all
scattering angles and for all product states, a situation that we term an
ideal resonance. By comparison, the limits of non-coherent control (i.e.
when no quantum interference occurs) are the cross sections corresponding
to the two initial states $\sigma_{q'}(11)$ and $\sigma_{q'}(22)$
(corresponding to $s=0$ and $s=1$).

Consider a system displaying an isolated Feshbach resonance \cite{Feshbach58}.
In this case the $T$ matrix elements at scattering energy $E$
are of the general form\cite{Feshbach58,Joachain83}
\begin{equation}
   \langle q',n |T| q,j \rangle =
      \gamma'_r(q',n)\gamma_r(q,j) / \left[ E-E_r \right] \;,
\label{res}
\end{equation}
where $\gamma'_r(q',n)$ and $\gamma_r(q,j)$ are matrix elements of the
Hamiltonian coupling the product state to the resonance state (denoted by
the subscript $r$) and coupling the resonance to the initial state,
respectively. $E_r$ is the complex energy associated with the resonance.
If the form of Eq.~(\ref{res}) holds, for all final states and all
scattering angles at the energy of interest, we have
\begin{eqnarray}
   \sigma_{q'}(kk)&=& { |\gamma_r(q,k)|^2 \over |E-E_r|^2 }
     \sum_n|\gamma'_r(q',n)|^2 \;, \;\;\;\; k=1,2 \nonumber \\
   \sigma_{q'}(12) &=& {\gamma^*_r(q,1)\gamma_r(q,2) \over
     |E-E_r|^2  } \sum_n |\gamma'_r(q',n)|^2 \;.
\label{partition}
\end{eqnarray}
so that
\begin{eqnarray}
&&\sigma_{q'} = {\sum_n|\gamma'_r(q',n)|^2 \over |E-E_r|^2 } \times
\nonumber \\ &&\left[ (1-s) |\gamma_r(q,1)|^2 + s |\gamma_r(q,2)|^2 +
2\sqrt{s(1-s)} |\gamma^*_r(q,1)\gamma_r(q,2)| \cos\left(
\mathrm{Arg}[\sigma_{q'}(12)] + \phi_{12} \right) \right].
\label{sigmares}
\end{eqnarray}
Under these circumstances the magnitudes of the $\sigma_{q'}(ij)$ satisfy
the Schwartz equality and complete control over $\sigma_{q'}$ is possible.

Control via a resonance over the {\em ratio} of cross sections into
different arrangement channels $q',q''$ requires that at least one of
these channels have a direct scattering component, or that they arise by
scattering through different ideal resonances. If this is not the case,
i.e. if both final arrangements $q'$ and $q''$ are accessible only via the
same ideal resonance, then, as is evident from Eq. (\ref{sigmares}),  the
ratio $\sigma_{q'}/\sigma_{q''}$ no longer depends upon the control
parameters $s$ and $\phi_{12}$.

The above argument is general and the recently observed reactive resonance
in $\rm{F+HD} \rightarrow \rm{D+HF}$ scattering
\cite{Castillo98,Skodje00,Schatz00} provides an opportunity for a
challenging numerical test within the framework of molecular scattering.
The FHD resonance, corresponding to a collinear arrangement with quantum
numbers $(v_{\mathrm FH}=3,v_{\mathrm HD}=0, v_{\mathrm bend}=0)$, is
short-lived, with a lifetime of 109 fs. In the low energy regime, where
the resonance dominates the reactive scattering, the rotational period is
$\approx 1420$ fs, and the contributions from several total angular
momentum values overlap one another \cite{Skodje00}). Advantageously, the
collinear FHD complex is much more likely to decay to $\mathrm{D+HF}$ than
to $\mathrm{H+DF}$, so that complete control over the resonant mechanism
should allow large control over both the $\mathrm{D+HF}$ cross sections as
well as over the ratio of $\mathrm{D+HF}$ to $\mathrm{H+DF}$ reactive
cross sections. Deviations of control from the maximum value can also be
be enlightening, being attributable to non-resonant, direct reactive
scattering contributions.

Control results shown below were obtained using a converged full
three-dimensional quantum coupled-channel hyperspherical coordinate
approach \cite{Skouteris00} on the Stark-Werner potential energy surface
\cite{Stark96}, using an available code\cite{code}. Results were obtained
for a variety of energies over the range 0.25--0.31~eV (relative to the HD
minimum) in increments of 0.005~eV. A basis set containing all three
channels with maximum diatomic energy levels of 1.7~eV, maximum rotational
quantum number $j_{\max}=15$, and maximum total angular momentum and
helicity $J_{\max}=31$ and $K_{\max}=4$ were used. $S$-matrix elements for
all possible combinations of energetically available initial and final
states were calculated.

Control was studied for the case where the initial state was comprised of
a superposition of the $(v_1,j_1,m_1) = (0,0,0)$ and $(v_2,j_2,m_2) =
(0,1,0)$ states of HD, which have internal energies of $e_1=0.23252$
and $e_2=0.24358$~eV, and with $k_1$ and $k_2$ satisfying the conditions
\cite{Brumer99}
\begin{equation}
  \mathbf{K}_1 = \mathbf{K}_2 \; ; \;\;\;\;
  E= \hbar^2k_1^2/2\mu + e_1 = \hbar^2k_2^2/2\mu + e_2.
\label{conditions}
\end{equation}
Here $\hbar\mathbf{K}_i$ and $\hbar\mathbf{k}_i$ are the center of mass
and relative momenta respectively, and $\mu=M_{\mathrm{F}}M_{\mathrm{HD}} /
M_{\mathrm{FHD}}$ is the reduced mass.  For the case we have considered the
kinetic energies of the two coherent components, defined as $E_{k_j} =
\hbar^2k_j^2/2\mu$  are related by $E_{k_2} = E_{k_1} - 0.01106$~eV.

Most relevant is the ratio of product cross sections. Figure~1a shows the
maximum and minimum values of the ratio of integral reactive cross
sections, $r = \sigma[\mathrm{D+HF}] / \sigma[\mathrm{H+DF}]$ attainable
at each energy.  Results using coherent control are shown as solid lines
while those without control [i.e. resulting from $\sigma_{q'}(11)$ and
$\sigma_{q'}(22)]$ are shown as dashed lines. Clearly, $r$ peaks near the
resonance ($E = 0.2550$~eV) with the ratio controllable, at resonance,
over a huge range: 2.60 to 161, a factor of 62. This range is far greater
than the analogous non-coherent control factor of 3.3. Further, the
coherent control parameters resulting in the minimum ratio (2.60) are
$s=0.7027$ and $\phi_{12}=185.1^\circ$, while for the maximum (161) they
are $s=0.9344$ and $\phi_{12}=352.9^\circ$, sufficiently far apart to
allow easy discrimination.

These results are shown in a somewhat different way in Figure 1b, where the
ratio of maximum and minimum ratios $r$, i.e. $R=\max[r] / \min[r]$
attainable via coherent control are compared to those attainable without
coherent control. Three features are evident: coherent control affords a
vastly larger range of control, the best ratio control is seen at the
resonance, and $R$ for the coherent control case remains greater that 20
at energies up to 0.31~eV, where the cross section is no longer dominated
by the resonant process. However, as seen in Fig 1a, in this higher energy
region the actual value of $r$ is quite small.

Figures 1c and 1d provide the cross sections from which the earlier panels
are constructed. Figure 1c shows that the $\mathrm{D+HF}$ cross section
can be coherently controlled at the resonance from a minimum of 0.0850
$\AA$ to a maximum of 2.193 $\AA$.  This is to be compared to the ideal
case of complete control with a minimum of 0 and a  maximum, corresponding
to $\sigma_{q'}(11)+\sigma_{q'}(22)$, of 2.278 $\AA$. Alternatively, we
can gauge the ``quality" of the resonance from the deviation of
$\sigma_{q'}(12)/[\sigma_{q'}(11) \sigma_{q'}(22)]^{1/2}$, from unity.
This ratio peaks, at the resonance energy, at a value of 0.906. As a
general observation we note that this ratio is far larger than that seen
in previous studies, such as those of the isotopic analogs of
$\mathrm{H+H_2}$ \cite{Brumer99}. Hence control is far more extensive in
this case than in any collisional process previously studied.

Results were also obtained for control of the differential cross section.
In this case the sum in Eq. (\ref{xs3}) does not contain an integral over
scattering angle. Results at the resonant energy show that the greatest
control occurs for backward scattering ($\theta=180^\circ$). There the
ratio of reactive cross sections can be coherently controlled from a
minimum of 0.236 (at $s=0.6469$ and $\phi_{12}=167.0^\circ$) to a maximum
of 1455 (at $s=0.8097$ and $\phi_{12}=331.1^\circ$), a factor of over
6000. By contrast, the non-coherent control factor never rises above 7 at
this energy.  At this scattering angle the Schwartz ratio
$\sigma_{q'}(12)/[\sigma_{q'}(11) \sigma_{q'}(22)]^{1/2} = 0.984$. Such
near-unity values are consistent with the fact that much of the averaging
is eliminated in examining the differential cross section.

We have shown above that if there is an ideal resonance then complete
control is possible using a superposition of two initial states. We note
further that the behavior of the Schwartz ratio at the resonance energy
provides a useful theoretical test for the contrapositive. That is, if the
Schwartz equality does not hold at a particular energy then the dynamics
must have a non-resonant component. Thus, for example, information
regarding the resonant (or non-resonant) behavior of H + HD scattering in
the forward direction, a subject of recent debate\cite{nature}, can be
augmented by computing the Schwartz ratio in that direction. Doing so
using the same initial superposition as for the F + HD above and the
established BKMP2 potential surface\cite{pes} gives us
$\sigma_{q'}(12)/[\sigma_{q'}(11) \sigma_{q'}(22)]^{1/2} =
0.850$, indicating that the dynamics does not go solely through a
resonance.

In conclusion, we have shown that scattering through a resonance allows
for a vast range of control over product cross sections using an initial
state comprised of a superposition of as few as two states. Preparing this
superposition for atom-diatom scattering does present a challenge since
the terms in the superposition are comprised of wavefunctions where the
translational and the internal diatom wavefunctions are
correlated\cite{Brumer99}. However, significant advances have been made in
experimental coherent control studies for unimolecular problems
\cite{CC1,CC2} and there are a number of possible approaches to preparing
the states required for bimolecular control. These include extending a
method successfully applied\cite{pritchard98} to atoms in order to prepare
the required correlated initial states, or state-selectively accelerating
or decelerating molecules\cite{meijer} that are in different internal
state to achieve this result. Alternatively, we note that, for the case of
the scattering of identical particles, preparation of the required states
is straightforward\cite{Gong03}. Since the discussion in this paper
applies equally well to such cases we suggest that the effect described
here may well be first observed, in molecular physics, in the scattering
of identical diatomic molecules, e.g. $\mathrm{OH + OH}$.

We thank Dr. D. Skouteris for sending us his auxiliary code for
calculating cross sections from the scattering output.  This work was
supported in part by the U.S. Office of Naval Research and by the Natural
Sciences and Engineering Research Council of Canada.

\pagebreak

\begin{figure}\caption{
Maximum and minimum values through which integral cross sections
can be coherently (solid line) or non-coherently (dashed line) controlled
using the initial state described in the text. (a) The ratio of the cross
sections into the $\mathrm{D+HF}$ channel to that in the $\mathrm{H+DF}$
channel. Note that the minimum values are very close to the $r=0$ line;
(b) The ratio of the maximum to minimum $r$ at each energy; (c) the cross
section for the resonant $\mathrm{D+HF}$ channel; (d) the cross section or
the non-resonant $\mathrm{H+DF}$ channel. }
\end{figure}




\begin{references}
\bibitem[*]{bylineone} Permanent address:  Chemical Physics Department,
 The Weizmann Institute of Science, Rehovot 76100, Israel.
\bibitem{CC1} M. Shapiro and P. Brumer, {\it Principles of the Quantum Control of
Molecular Processes} (Wiley, New York, 2003); M. Shapiro and P. Brumer,
Adv. At. Mol. Opt. Phys. {\bf 42}, 287 (2000).
\bibitem{CC2}  S.A. Rice and M. Zhao, {\it Optical Control of Molecular Dynamics}
(Wiley, New York, 2000); S.A. Rice, Nature {\bf 409}, 422 (2001).
\bibitem{Shapiro96} M. Shapiro and P. Brumer, Phys. Rev. Lett. {\bf 77}, 2574
(1996); A. Abrashkevich, M. Shapiro and P. Brumer,
    Phys. Rev. Lett. {\bf 81}, 3789 (1998); {\it erratum} {\bf 82}, 3002 (1999);
    Chem Phys. {\bf 267}, 81 (2001).
\bibitem{Brumer99} P. Brumer, A. Abrashkevich and M. Shapiro,
    Discuss. Faraday Soc. {\bf 113}, 291 (1999).
\bibitem{Frishman99a} E. Frishman, M. Shapiro and P. Brumer, J. Chem. Phys.
    {\bf 110}, 9 (1999). Note that the result obtained in this reference applies
separately to each total angular momentum contributing to the scattering.
(See, Ref. \cite{CC1}).
\bibitem{Gong03} P. Brumer, K. Bergmann and M. Shapiro, J. Chem.
Phys.{\bf 113}, 2053 (2000); J. Gong, M. Shapiro and P. Brumer, J. Chem.
Phys. {\bf 118}, 2626 (2003).
\bibitem{Castillo98} J.F. Castillo and D.E. Manolopoulos,
    Discuss. Faraday Soc. {\bf 110}, 119 (1998).
\bibitem{Skodje00} R.T. Skodje, D. Skouteris, D.E. Manolopoulos,
    S.-H. Lee, F. Dong and K. Liu, J. Chem. Phys. {\bf 112}, 4536 (2000);
 Phys. Rev. Lett. {\bf 85}, 1206 (2000).
\bibitem{Schatz00} G.C. Schatz, Science {\bf 288}, 1599 (2000);
W.W. Harper, S.A. Nizkorodov and D.J. Nesbitt,
    J. Chem. Phys. {\bf 116}, 5622 (2002);
 S.-H. Lee, F. Dong and K. Liu, J. Chem. Phys. {\bf 116},
    7839 (2002).
\bibitem{Feshbach58} H. Feshbach, Ann. Phys. (N.Y.) {\bf 5}, 357 (1958);
    {\bf 19}, 287 (1962).
\bibitem{Joachain83}  C.J. Joachain, {\it Quantum Collision Theory},
    3rd ed. (American Elsievier Pub. Co., New York, 1983).
\bibitem{Skouteris00} D. Skouteris, J.F. Castillo and D.E. Manolopoulos,
    Comp. Phys. Comm. {\bf 133}, 128 (2000).
\bibitem{Stark96} K. Stark and H.-J. Werner, J. Chem. Phys. {\bf 104}, 6515 (1996).
\bibitem{code} The code is available at www.cpc.cs.qub.ac.uk/cpc/.
\bibitem{nature} S.C. Althorpe et al, Nature {\bf 416}, 67 (2002); S.A. Hatich
et al, {\em ibid.} {\bf 419}, 281 (2002).
\bibitem{pes} A.I. Boothroyd et al, J. Chem. Phys. {\bf 104}, 6515 (1996)
\bibitem{pritchard98} E.T. Smith, et al, Phys. Rev. Lett. 81, 1996 (1998)
\bibitem{meijer} H.A. Bethlem et al, Phys. Rev. Lett. 88, 133003 (2002)
\end{references}
\end{document}